# Time Segmentation Approach Allowing QoS and Energy Saving for Wireless Sensor Networks

Gérard Chalhoub, François Delobel, and Michel Misson

**Abstract**—Wireless sensor networks are conceived to monitor a certain application or physical phenomena and are supposed to function for several years without any human intervention for maintenance. Thus, the main issue in sensor networks is often to extend the lifetime of the network by reducing energy consumption. On the other hand, some applications have high priority traffic that needs to be transferred within a bounded end-to-end delay while maintaining an energy efficient behavior. We propose MaCARI, a time segmentation protocol that saves energy, improves the overall performance of the network and enables quality of service in terms of guaranteed access to the medium and end-to-end delays. This time segmentation is achieved by synchronizing the activity of nodes using a tree-based beacon propagation and allocating activity periods for each cluster of nodes. The tree-based topology is inspired from the cluster-tree proposed by the ZigBee standard. The efficiency of our protocol is proven analytically, by simulation and through real testbed measurements.

———————————— ◆ ————————————

## 1 INTRODUCTION

WIRELESS sensor networks are recently highly considered for deployment in various domains such as military, health, transportation, industrial, etc. They are constituted of communicating nodes that exchange data in an ad hoc manner to relay information such as temperature, pressure or movement observed in the coverage area of the sensor device. Nodes usually run on batteries which makes the energy efficiency an essential concern for communication protocols. On the other hand, some applications demand a certain quality of service mainly in terms of end-to-end delay. Critical data must be delivered to the supervising unit within a certain bounded delay in order to be able to take appropriate action in time.

In most deployment cases, wireless sensor networks topologies spread over a multi-hop configuration and data should travel in a hop-by-hop manner to reach the sink. In order to offer a bounded end-to-end delay, access to the medium should be guaranteed or time bounded on each hop. In this paper we present MaCARI an energy efficient MAC layer protocol based on a time segmentation approach with guaranteed access to the medium offering a bounded end-to-end delay for high priority traffic. We demonstrate how this segmentation approach enhances the overall network performance and how bounded end-to-end delay affects the network size. Our work is based on testbed measurements and simulations.

The paper is structured as follows. In section 2 we present an overview of the different MAC protocols proposed for wireless sensor networks. In section 3 we describe the time segmentation approach of our protocol. In section 4 we evaluate the network performance in terms of packet loss rate and collisions, we evaluate as well the energy efficiency of MaCARI. In section 5 we evaluate the maximum network size that guarantees a certain end-to-end delay constraint. We finally conclude in section 6.

## 2 STATE OF THE ART

A large number of energy-efficient MAC protocols have been proposed for wireless sensor networks. Most of these protocols try to avoid the essential sources of energy consumption at the MAC level, which are: *overhearing* that is the reception of unwanted traffic, *collisions* that is the simultaneous reception of more than one frame, *idle listening* that is being active without neither receiving nor transmitting, and *overhead* that is the exchange of non application related data. These protocols can be categorized into contention-based protocols, TDMA-based protocols and hybrid protocols that are based on both.

### 2.1 Contention-based protocols

Contention-based protocols are based on the use of CSMA/CA algorithm to avoid collisions.

One of the first energy efficient MAC protocol based on contention is PAMAS [1] PAMAS is designed to reduce overhearings by using a separate channel to exchange RTS/CTS messages endicating the duration of the communication that is going to take place. S-MAC [2] adopts the same principle as PAMAS by using only one

————————————————

• *Gérard Chalhoub, François Delobel and Michel Misson are with Clermont University/ LIMOS-CNRS, Complexe scientifique des Cézeaux, 63177 Aubière cedex, France.*

channel. Moreover, S-MAC reduces the idle listening by allowing nodes to switch off periodically their radio devices during predefined time intervals. T-MAC [3] is an improvement of S-MAC that dynamically reduces the active period when no activity is detected in order to sleep more and save energy more often. D-MAC [4] on the other hand was proposed to reduce the latency for message forwarding by preventing intermediate nodes between the source and the destination to go to sleep.

These protocols offer a flexible, decentralized and energy-efficient behavior, but fail to guarantee access to the medium and hence do not offer a bounded end-to-end delay.

## 2.2 TDMA-based protocols

In TDMA-based protocols, nodes are allocated time slots to exchange messages with no risk of collisions.

TRAMA [5] is considered one of the first MAC protocols based on TDMA that allow energy saving by putting nodes to sleep mode when there are neither transmitter nor receiver during a given time slot. FLAMA [6] improves TRAMA by using a more simple algorithm for time slot allocations. ER-MAC [7] and DE-MAC [8] allow critical nodes (nodes with limited remaining energy level) to sleep longer.

TDMA-based protocols offer guaranteed access to the medium but are based on complex algorithms that overload the network.

## 2.3 Hybrid protocols

Hybrid protocols alternate the nodes activity between periods of TDMA mode and periods of CSMA/CA mode.

Funneling-MAC [9] and G-MAC [10] allocate additional time slots for nodes that are close to a central node to exchange messages in TDMA mode. In Z-MAC [11] the TDMA slots are not guaranteed for their owners, other nodes can use them as well. Owners have a priority only during the start of the slot. The IEEE 802.15.4 standard [12] defines a MAC protocol that saves energy by putting nodes periodically in sleep mode and offers guaranteed access to the medium during specific time slots called GTS (Guaranteed Time Slot). The ZigBee standard [13] is based on the IEEE 802.15.4 but fails to keep the guaranteed aspect of the GTS and does not offer a guaranteed end-to-end delay.

## 3 MACARI PROTOCOL

In this section we describe the time segmentation on which the MaCARI protocol is based and explain how using guaranteed time intervals can achieve a bounded and-to-end delay.

### 3.1 Time segmentation of MaCARI

MaCARI is based on a cluster-tree topology, similar to the one proposed by ZigBee. Clustering techniques have been considered as an efficient way to consume less energy as discussed in [14], [15]. The network supports three types of nodes: the *PAN coordinator* (Personal Area Network coordinator) which is the first node to be activated in the network, coordinators which are nodes that allow other nodes to join the network, and end-devices nodes which are attached to only one coordinator and can only communicate to it. We call a star the coordinator and the end-devices associated to it. The coordinators (the PAN coordinator included) allocate hierarchical addresses to the nodes according to three topology parameters which are the *Rm* (maximum number of coordinator children), *Cm* (maximum number of children) and *Lm* (maximum depth of the tree). Frames are relayed along the cluster-tree from node to node and next hops are chosen based on these hierarchical addresses.

The PAN coordinator is informed of each new coordinator that joins the network. Based on this information, the PAN coordinator defines a global cycle as illustrated on Fig. 1. It allocates for each star a time interval during which the coordinator communicates with its end-devices. Each star gets an activity period as shown on Fig. 2. To inform all the nodes of the network about the activity periods that were allocated, the PAN coordinator generates a beacon frame containing the necessary information. This beacon frame is propagated along the tree in a certain order specified inside the beacon frame to avoid collisions. The time interval needed to propagate the beacon frame is called the synchronization period (interval [T0; T1] of Fig. 1). At the end of this period, all the nodes are synchronized according to the sequence of the activity periods specified by the PAN coordinator (interval [T1; T2] of Fig. 1). Devices of a star are not allowed to transmit during the activity period of other stars.

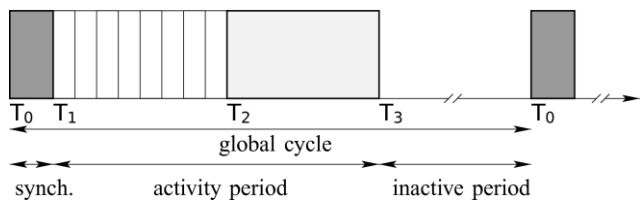

Fig. 1. The global cycle of MaCARI.

The reservation of an activity period for each star during [T1; T2] makes it possible for a coordinator to allocate GTS for its end-devices with no risk of overlapping GTS allocation between different stars. These GTS will be used to exchange high priority traffic between end-devices and coordinators. Since traffic generated in each star is usually destined to the sink, high priority frames should travel in a hop-by-hop matter along the tree to reach the sink. In what follows, we explain how high priority traffic is exchanged between coordinators.

### 3.2 Guaranteed end-to-end delay

To take into account the high priority traffic generated in each star, guaranteed time slots are allocated for each coordinator to communicate with its parent coordinator. These slots are called *relay intervals* and they come at the end of the activity period of each star as shown on Fig. 2. On the other hand, low priority traffic is exchanged between coordinators during [T2; T3] using CSMA/CA and a traditional routing protocol like OLSR [16] or AODV [17]. We do not evaluate the activity of the network during [T2; T3] in this paper.

By activating the stars starting from the bottom of the tree, the activity periods sequence allows a frame to go up the tree from the source to the sink (the PAN coordinator) in one global cycle. In this case, the activity periods are scheduled in an up-stream manner. We suppose that the relay intervals were allocated taking in consideration the size of the high priority traffic generated and accumulated at each level of the tree.

According to the example given on Fig. 2, if a frame destined to A is generated by an end-device associated to G, it is sent to G during the activity period of star G, the frame is then relayed during the relay interval between G and C. C then relays this frame to A during the relay interval between C and A.

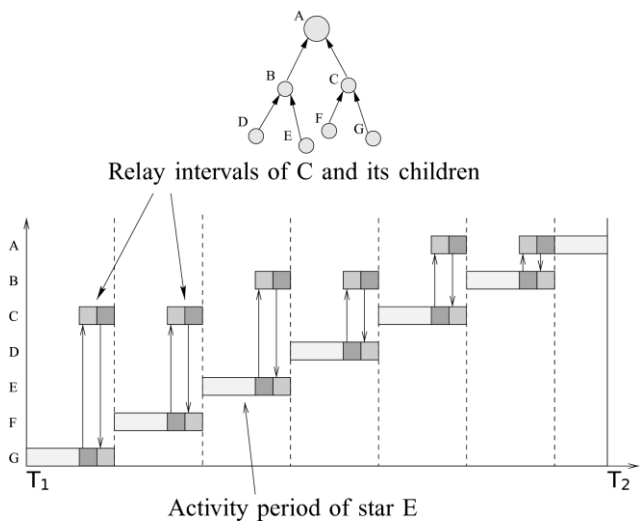

Fig. 2. MaCARI allocates an activity period for each star during [T1; T2].

In a more general configuration, where the final destination for the high priority traffic can be any node of the network, the activity periods can be scheduled alternatively in an up-stream or a down-stream manner. Thus, prioritizing an up-stream flow (from the end-devices to the sink) for the high priority traffic during one global cycle and prioritizing a down-stream flow (from the sink to the end-devices) in the next global cycle. In this paper we consider that the up-stream flow is the more common scenario and we consider our evaluation based on that assumption.

## 4 OVERALL EVALUATIONS

We have implemented MaCARI on the NS2 simulator to evaluate its performances and on the B2400ZB-tiny network cards to prove its feasibility. In this section we present simulation results for the overall netwrok performance in terms of number of collisions and medium access efficiency.

### 4.1 Simulation Results

In this section, we evaluate our protocol MaCARI through simulation by comparing it, on one hand, to a cluster-tree configuration that uses a beacon-only period approach to avoid beacon frames collision (this configuration is depicted on Fig. 3), and on the other hand, to a version of MaCARI without the relay intervals between parent/child coordinators to emphasize on the benefits of these guaranteed time intervals.

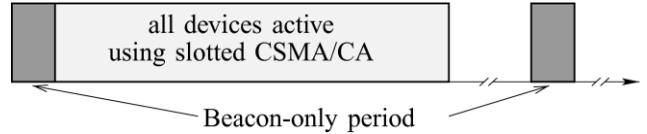

Fig. 3. Beacon-only period approach configuration.

All the simulations performed in this paper are based on the same simulation setup, with varying parameters. We chose five different scenarios given on table 1. These scenarios represent different sizes for industrial wireless sensor networks. The simulator used is Network Simulator 2 (version 2.31). We implemented the IEEE 802.15.4 physical layer by taking into account the capture effect.

TABLE 1
SIMULATION SCENARIOS

| scenario | nb. of coordinators | nb. of end-devices | Cm |
|---|---|---|---|
| 1 | 9 | 25 | 6 |
| 2 | 9 | 36 | 7 |
| 3 | 16 | 49 | 6 |
| 4 | 16 | 64 | 7 |
| 5 | 25 | 81 | 7 |

Traffic is destined to the PAN coordinator. Each end-device generates 16 frames, with a frequency of one frame per second. We consider that the network is already created and nodes are already associated. Rm is fixed to 3 and Lm is fixed to 5 for all the scenarios. We allocated for each star the same activity period duration. This duration is enough to collect one frame per end-device. At the end of each simulation, we compute the overall number of received frames, the overall number of sent frames and the overall number of collisions. Notice that we do not include the number of retransmissions in case of failures to access the medium.

The routing algorithm used in [T1; T2] during the relay intervals is the hierarchical routing algorithm used in ZigBee [13]. The routing algorithm used in [T2; T3] is a modified hierarchical routing algorithm which uses a table of neighbors to determine if a neighbor can reach the final destination of the packet with less hops than the default tree route. This modified hierarchical routing algorithm is similar to the Shortcut Tree Routing protocol described in [18]. We implemented the ITU [19] propagation model for more realistic propagation model for indoor communications.

The duration of the activity period of each star is fixed to 61,44ms. This time is chosen according to the results of our prototype [20] (it takes up to 61.44ms for 8 end-devices to send one data frame to their coordinator with slotted CSMA/CA). We considered that the duration of [T2; T3] is equal to the duration of [T1; T2]. So, in the case of a cluster-tree configuration with a beacon-only period approach, the activity period must be equal to [T1; T2] + [T2; T3]. We fixed the duration of the relay interval to 15.36ms taken from [T2; T3]. Note that the overall time allocated for both relaying during [T1; T2] and routing during [T2; T3] stays invariant. This choice is driven by the fact that [T2; T3] and the relay intervals of [T1; T2] are both dedicated to routing activities. We considered that 25% of the generated traffic is a high priority traffic that has to be relayed in the relay intervals.

The graphs are the average of 10 replications over 10 different topologies for each scenario. This gives us an average of 100 replications for each value given for each scenario. Fig. 4 shows how using the time segmentation approach increases the overall number of received frames in the network respectively. It is interesting to note on Fig. 4 that for approximately 55Kbytes sent, only 25Kbytes are received in the cluster-tree configuration without the time segmentation during [T1; T2], and 50Kbytes are received by adopting the time segmentation. This is essentially due to the fact that by activating the nodes sequentially by small groups (by stars in our case) instead of activating all the nodes during the same period of time is more efficient for the slotted CSMA/CA.

On the other hand we notice that relay intervals achieve a significant reduction in the number of collisions in all scenarios as shown on Fig. 5, this is essentially due to the fact that 25% of the traffic is sent during the relay intervals where there is no risk of collisions.

## 4.2 Energy Efficiency

In order to evaluate the energy efficiency of MaCARI, we evaluated the duration of each interval according to the number $n$ of coordinators and $m$ end-devices. We estimate that when nodes are active they consume the same amount of energy for whatever state they are in: idle, receive or send. Thus, we compute the ratio of the time spent in active mode compared to the time spent in inactive mode. Thus the total energy spent in the network is the sum of the time spent in active state for all the nodes of the network. Let $E$ be the total energy spent when the nodes are active all the time during a global cycle.

$$E = (m+n) \times D \quad (1)$$

Where $D$ is the duration of the global cycle.

We considered the active duration for the cluster-tree configuration with the beacon-only period approach without time segmentation equal to the duration of [T1; T3]. Thus, we obtain the energy $E'$ spent during one cycle for the beacon-only period approach:

$$E' = (m+n) \times D_{T0T3} \quad (2)$$

Where $D_{T0T3}$ is the duration of [T0; T3].

For MaCARI, by considering the duration of the relay interval as one third of the activity period of the interval allocated for a coordinator, we have:

$$E'' = (m+n) \times D_{T0T1} + n \times ((D_{T1T2}/n) + (1/3 \times Rm \times D_{T1T2}/n) + D_{T2T3}) \quad (3)$$

Where $D_{T1T2}/n$ denotes the duration of the activity period allocated for each star (including the relay interval) supposing that all stars are allocated the same duration.

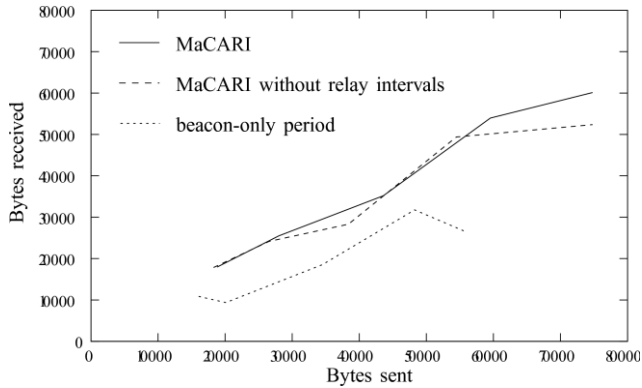

Fig. 4. MaCARI increases the medium access control efficiency.

Fig. 6 shows the energy gain that MaCARI offers compared to a configuration with cluster-tree configuration without the time segmentation. We varied the length of the inactivity period [T3; T0] so that it takes the following values: 0, $D_{T1T3}/2$, $D_{T1T3}$ and $2*D_{T1T3}$ ($D_{T1T3}$ denotes the duration of [T1; T3]). We fixed the relay intervals to 20ms and we considered $Rm = 3$ for all the scenarios ($Cm$ is given on table I). It is interesting to see that the energy consumption in all the scenarios are similar, which explains why the curves are close to one another.

## 5 NETWORK DIMENSIONING

In this section we explain the frame deference phenomenon on which we base our end-to-end delay evaluation. Then, we evaluate the network size in terms of number of devices (coordinators and end-devices) that enables us to guarantee a bounded end-to-end delay.

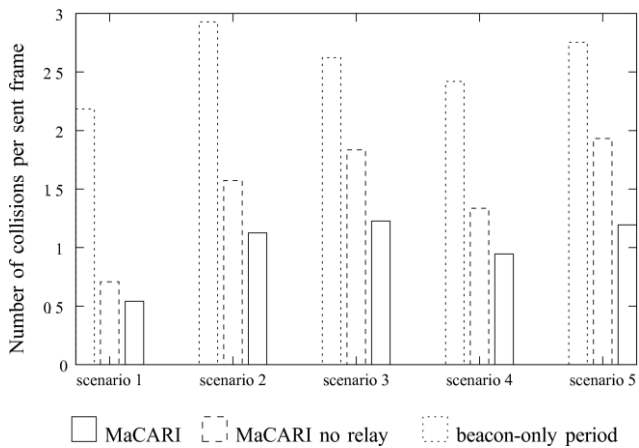

Fig. 5. MaCARI decreases the number of collisions.

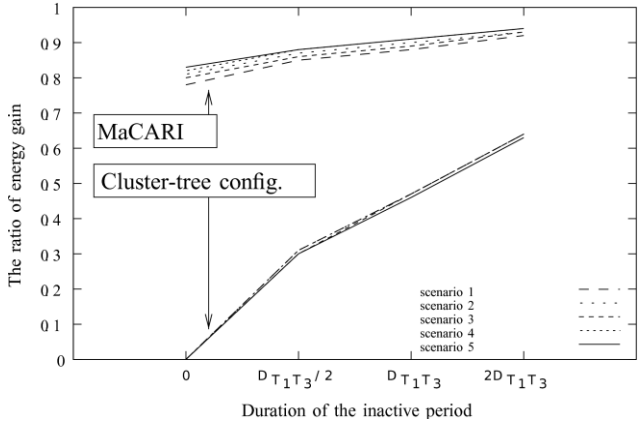

Fig. 6. Energy efficiency of MaCARI.

## 5.1 Frame Deference

When a MAC layer has a bounded time interval during which it is allowed to transmit, frames generated outside that time interval are kept in queues until reaching the start of that interval. In addition, to prevent nodes from causing an overflow by transmitting when the interval has ended, frames that are generated towards the end of the time interval are not transmitted by the MAC layer if the time left in the time interval is not enough to finish the transmission transaction (the frame transmission and the MAC level acknowledgment transmission). This phenomenon is called *frame deference* and was mentioned in [21] where authors pointed at the collisions it caused in the 2003 version of IEEE 802.15.4 [22].

In the case of MaCARI, we will consider the frame deference of the high priority traffic. This traffic is generated during the GTS of an end-device and then relayed along the tree during the relay intervals. In what follows, we will consider the wost case scenario of the end-to-end delay that might take place and evaluate the network size according to that.

## 5.2 Worst Case Scenario for End-to-End Delay

The worst case scenario occurs when the frame reaches the MAC layer δ seconds before the end of the GTS of the end-device, and this δ time is not enough to achieve the transmission transaction. Fig. 7 depicts the δ time interval according to the global cycle and the activity period. In that case, the frame transmission is deferred until the next activity period of the next global cycle. Thus, the delay induced by this deference is at most equal to the duration of a global cycle.

The end-to-end delay is affected by the access delay which is the time spent by the MAC layer to access the medium including the frame deference, and the number of hops that separate the source from the destination. In the case of high priority traffic which is exchanged during the guaranteed relay intervals, the access delay is deterministic. Indeed, the frames are sent directly to the medium without any backoff delays. Thus the access delay is limited to the frame deference delay.

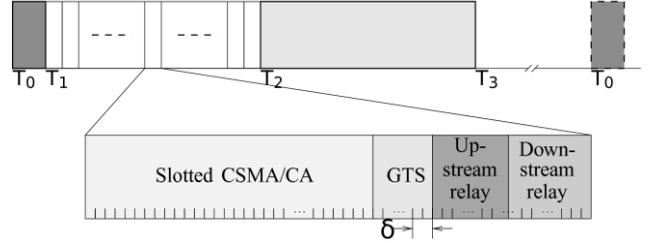

Fig. 7. Worst case scenario for end-to-end delay.

If we adopt an up-stream scheduling for the activity periods during [T1; T2], the frame can reach the sink in one global cycle (more precisely in one [T1; T2] interval) if it is not deferred. In case the frame is deferred, it has to wait for the activity period of the next global cycle. Hence, by considering the worst case scenario, the end-toend delay does not exceed the duration of a global cycle plus one [T1; T2] interval. Which gives us the following equation in case the maximum end-to-end delay tolerated is 1 second:

$$D_{T0T1} + D_{T1T2} + D_{T2T3} + D_{T3T0} + D_{T1T2} < 1 \quad (4)$$

## 5.3 Duration of Time Intervals

We evaluated the duration of each interval on testbeds using the B2400ZB-tiny motes in terms of number of coordinators.

The duration of [T0; T1] is a function of the number of coordinators and it was evaluated in [23]. Based on real measurements using a logical state analyzer, we calculated the time needed to process, prepare and send the beacon frame. We obtained the following equation:

$$D_{T0T1} = n \times (0.00032 \times n + 0.008) \quad (5)$$

Where $n$ is the number of coordinators, 8ms is the time needed for processing beacon frame when we have only one coordinator in the network (it is the case when the PAN coordinator is the only entity in the network), 0.32ms is the time needed to take into account each new coordinator that is added to the beacon.

On the other hand, each new coordinator joining the network is added to the beacon frame and this affects both the transmission duration and the processing time of the beacon frame which explains the quadratic form.

As for the duration of [T1; T2], it is a function of the duration needed for each coordinator to collect the data generated by its end-devices. We made the assumption that all the stars have the same number of active end-devices during a given global cycle and that this number stays the same for the duration of the activity period.

We evaluated the time needed for a coordinator to collect one frame from each of its active end-devices using slotted CSMA/CA of IEEE 802.15.4. We considered 4 representative sizes for the stars: 2, 4, 6 and 8 active end-devices. Each end-device has one frame to transmit at the start of the activity period. We calculate the duration that separates the start of the actity period and the instant the last frame was received by the coordinator. The results given on table 2 are based on testbed measurements over 75 replicas by considering the upper bound of the maximum value obtained for each size.

TABLE 2
THE DURATION OF THE ACTIVITY PERIOD IN MS NEEDED TO SUCCESSFULLY TRANSMIT 1 FRAME PER END-DEVICE

| Number of active end-devices | 2 | 4 | 6 | 8 |
|---|---|---|---|---|
| Collect duration in ms | 20 | 30 | 45 | 50 |

It is interesting to notice that the collect duration increase in a logarithmic manner over the number of active end-devices in the star. Let us now consider a scenario where each coordinator has 4 active end-devices in each global cycle (which gives us 30ms for each coordinator to collect the data), and by considering a relay interval of 10 ms, we obtain the following equation:

$$D_{T1T2} = (0.03 + 0.01) \times n \quad (6)$$

As our objective is to obtain a maximum delay of 1 second, we can consider that there's no inactive period to make the global cycle shorter. Even though [T3; T0] is null, MaCARI can still achieve energy efficiency by reducing the activity duration by 80% during a global cycle, as shown on Fig. 6.

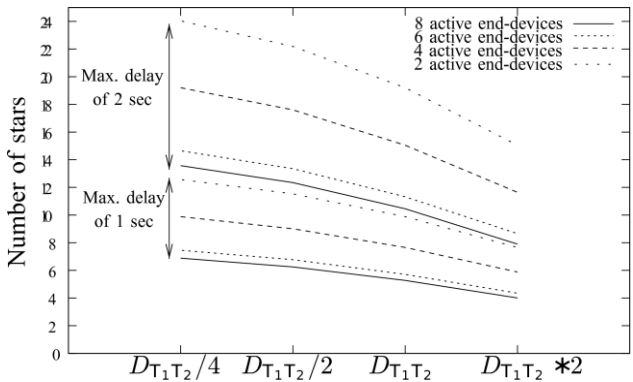

Fig. 8 Maximum network size in terms on number of stars that guarantees an end-to-end delay of 1 second and 2 seconds.

By applying equations (2) et (3) to equation (1), and by considering [T2; T3] = [T1; T2], we obtain an equation with only one variable, which is the number of coordinators, which in fact represents the number of stars of a certain size in terms of number of end-devices (4 active end-devices per star in this example).

This second degree equation gives us 1 positive value for $n$ which is 7.8. Thus, in order to guarantee a bounded end-to-end delay from any source to the sink of 1 second, the network size should not exceed 7 stars with 4 active end-devices each during a given global cycle.

We varied the duration of [T2; T3] from $D_{T1T2}/4$ to $D_{T1T2} * 2$, where $D_{T1T2}$ is the duration of [T1; T2] and this for a bounded end-to-end delay of 1 second and 2 seconds. Results in terms of number of stars are given on Fig. 8.

In all the scenarios considered in Fig. 8 we only consider the active end-devices of a star. Thus, the size of the network represents the active part of the network. A star can maintain 16 end-devices with only half of them active during a given global cycle, and the other half active in the next global cycle.

If we take the example with 4 active end-devices per star, with $D_{T2T3} = D_{T1T2}/4$, in order to obtain a maximum end-to-end delay of 2 seconds, the active part of the network in a given global cycle should not exceed 80 end-devices and 20 coordinators. In a practical real case scenario, it can represent a wireless sensor network of 240 end-devices with a periodic traffic generation of 1 frame every 6 seconds. This gives us, with a uniform asynchronous periodic traffic generation by the end-devices, a traffic generation of 80 frames every 2 seconds (a similar configuration of 80 active end-devices).

## 5.4 End-to-End Delay Evaluation

In this section we consider 2 representative scenarios in order to validate the results obtained in Fig. 8. The first scenario has 5 stars and 8 active end-devices per star, which represents a dense configuration, and the second scenario has 9 stars and 2 active end-devices per star, which represents a less dense but larger configuration. We considered the topologies depicted on Fig. 9 and Fig. 11 as examples for the 2 scenarios respectively where we considered evaluating the end-to-end delay for the traffic generated by end-devices 44 and 26 respectively.

We simulated using NS2 version 2.31 the two scenarios. We included a timestamp indicating the instant the frame was generated in terms of backoff periods: Frames are timestamped with the number of backoff periods spent since the instant T1 of a given global cycle. Global cycles are identified using a sequence number included in the beacon frame. Using this time reference we are able to calculate the end-to-end delay by comparing the instant the frame is received by the final destination to the timestamp of the frame.

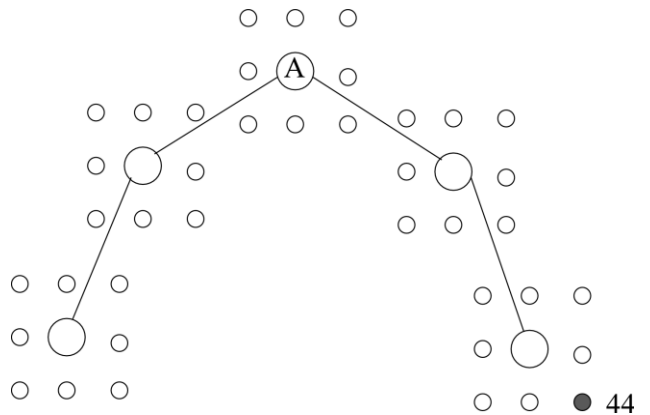

Fig. 9. A topology with 5 coordinators and 8 active end-devices per coordinator.

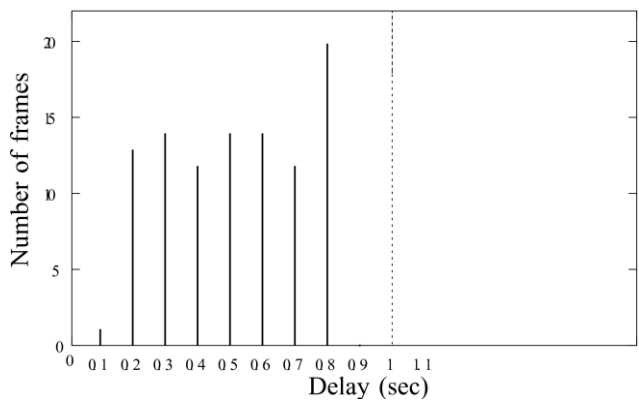

Fig. 10. End-to-end delay for the traffic sent by end-device 44 of Fig. 9.

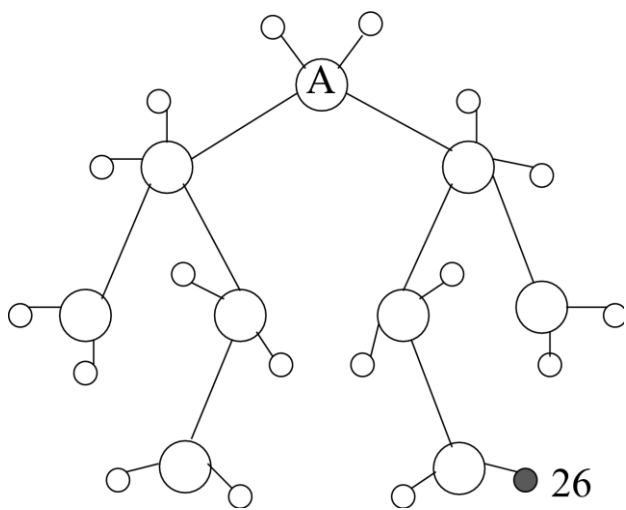

Fig. 11. A topology with 9 coordinators and 2 active end-devices per coordinator.

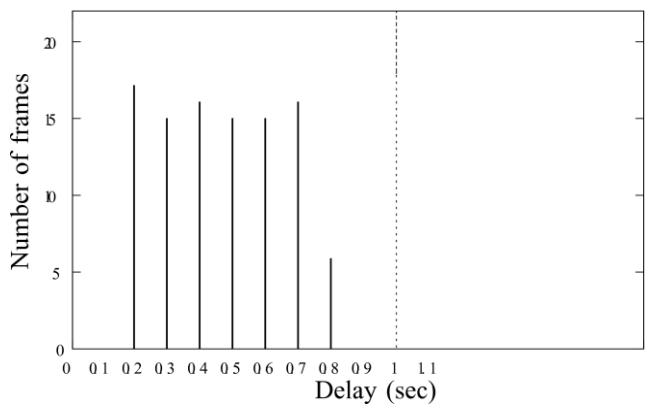

Fig. 12. End-to-end delay for the traffic sent by end-device 26 of Fig. 11.

We considered 100 frames destined to the sink. We considered end-device 44 on Fig. 9 and end-device 26 on Fig. 11 as source of traffic. Results on Fig. 10 and Fig. 12 show the end-to-end delay for the 100 frames generated. For the configuration on Fig. 9, the duration of the global cycle is 748 ms. We can notice how 32 frames had an end-to-end delay between 0.7 and 0.9 second, which means that these 32 were generated towards the end of the GTS and deferred to the next global cycle. For the configuration on Fig. 11, the duration of the global cycle is 491,68ms. In this configuration, we obtained 52 deferred frames.

It is interesting to notice that all the 100 frames in both configurations have been received. Thus, 100 % of the high priority traffic generated was received in less than 1 second.

## 7 CONCLUSION

Wireless sensor networks are known for energy efficient behavior that increases the network lifetime. These networks are being deployed in many industrial applications today where the main concern is not only the network lifetime but also the quality of service especially in terms of end-to-end delay. In this paper, we presented how the MaCARI protocol is able to achieve a guaranteed end-to-end delay for high priority traffic while maintaining an overall energy efficient behavior. We proved, based on simulations using NS2, how MaCARI improves the access to the medium by increasing the quantity of traffic successfully received and by decreasing the number of collisions compared to a beacon-only period approach of the cluster-tree.

We evaluated the network size in terms of number of active devices based on testbed measurements and simulations: we pointed to the fact that in order to guarantee a bounded end-to-end delay, the network size is also bounded. Other configurations guaranteeing even smaller delays for a higher number of devices in the network are possible by activating stars simultaneously using different frequencies for each star. Using a different frequency in each star reduces significantly the duration of the global cycle of MaCARI and thus the end-to-end delay.

In the future, an adaptive time interval allocation will be considered. For example, the number of descendants should be taken into account for allocating the relay intervals.

## ACKNOWLEDGMENT

This work was partially funded by the ANR OCARI project [24].

**Gérard Chalhoub** obtained his Masters and his PhD in Computer Science in 2006 and 2009 respectively at Blaise Pascal University, Clermont-Ferrand in France. He is currently a temporary teaching assistant at the Networks and Telecommunications Department and researcher in the research team "Réseaux et Protocoles" of Computer Science Laboratory of Clermont University: LIMOS-CNRS. His current research interests are wireless sensor networks protocols engineering.

**François Delobel** obtained his PhD in Computer Science in 2000 in the University of Nice Sophia Antipolis, France. After being a temporary teaching assistant in the University of Lyon, he joined the departement of Computer Science of the Institute of Technology in the University of Clermont-Ferrand. He is now teaching Object Programming and System Programming & Administration as a lecturer and works in the research team "Réseaux et Protocoles" of the laboratory LIMOS-CNRS. His research interests moved from Artificial Intelligence to wireless sensor networks protocols engineering and embedded systems.

**Michel Misson** obtained his PhD thesis in Nuclear Physics and his "Habilitation à Diriger des Recherches" (HDR) degree respectively in 1979 and 2001 both at Blaise Pascal University, Clermont-Ferrand in France. In 1983, he became a lecturer, teaching networks and system architecture in the Computer Science Department of the Institue of Technology in Clermont-Ferrand. He is now Professor in the Networks and Telecommunications Department and he manages the research team "Réseaux et Protocoles" of the Computer Science Laboratory of Clermont University: LIMOS-CNRS. He served on many conference committees and journals reviewing processes. He has more than 50 publications. His current research interests are Wireless Local Area Networks, Low Power Wireless Personal Networks, Wireless Sensor Networks, real-time systems and protocol engineering.